\documentclass[12pt,preprint]{aastex}
\def \kms {km s$^{-1}$ }
\def \lsol {L$_{\odot}$ }
\def \msol {M$_{\odot}$ }
\def \kmss {km s$^{-1}$}

\def \msols {M$_{\odot}$}

\slugcomment{The Astrophysical Journal, May 20, 2007, Volume 661}
\lefthead{Baan et al.}
\righthead{HI and OH Absorption toward NGC$\,6240$}
\begin{document}
   \title{HI AND OH ABSORPTION TOWARD NGC$\,6240$}
\author{Willem A. Baan}
\affil{ASTRON, Westerbork Observatory\footnote{The Westerbork
Observatory is part of the Netherlands Foundation for Research in
Astronomy (NFRA-ASTRON) and is partially funded by the
Organization for Scientific Research (NWO) of The Netherlands.},
            P.O. Box 2, 7990 AA Dwingeloo, The Netherlands}
\email{baan@astron.nl}
\author{Yoshiaki Hagiwara}
\affil{National Astronomical Observatory of Japan,
            2-21-1 Osawa, Mitaka,Tokyo, Japan}
\email{yoshiaki.hagiwara@nao.ac.jp}
\author{Peter Hofner}
\affil{Physics Department, New Mexico Tech, Socorro, NM 87801\\
and\\
National Radio Astronomy Observatory\footnote{The National Radio
Astronomy Observatory is a facility of the National Science
Foundation operated under cooperative agreement by Associated
Universities, Inc.}, P.O. Box O, Socorro, NM 87801}
\email{phofner@nrao.edu}
%
%

%
\begin{abstract}
{VLA observations of large-scale HI and OH absorption in the
merging galaxy of NGC$\,$6240 are presented with 1 arcsec
resolution. HI absorption is found across large areas of the
extended radio continuum structure with a strong concentration
towards the double-nucleus. The OH absorption is confined to the
nuclear region. The HI and OH observations identify fractions of
the gas disks of the two galaxies and confirm the presence of
central gas accumulation between the nuclei. The data clearly
identify the nucleus of the southern galaxy as the origin of the
symmetric superwind outflow and also reveal blue-shifted
components resulting from a nuclear starburst. Various absorption
components are associated with large-scale dynamics of the system
including a foreground dust lane crossing the radio structure in
the northwest region.}
\end{abstract}
%
%
%
%
\section{Introduction}

The chaotic appearance of NGC$\,6240$ at all wavelengths is due to
a forceful galactic collision of two galaxies (Fosbury \& Wall
1979). The two individual nuclei of NGC$\,6240$ were first
detected in R and I bands at a projected distance of 1.8" or 0.9
kpc (Fried \& Schulz 1983). Early H$\alpha$ studies revealed
extended emission with two independent and almost perpendicular
disk systems (Bland-Hawthorn et al. 1991).

NGC$\,6240$ is a prototypical Luminous Infrared Galaxy with an IR
luminosity of L$_{IR}$ = L$_{8-1000 \mu m}$ = 6 x 10$^{11}$ \lsol
(Sanders et al. 1988). The FIR luminosity of these galaxies is
powered by extremely high star-formation activity and or an
embedded AGN. For NGC\,6240 the mid-IR observations are consistent
with a dominant starburst power contribution of approximately
75$\%$ within the central 5 kpc (Genzel et al. 1998).

Radio data show two nuclei embedded in a connecting structure that
extends into a loop structure to the West (Condon et al. 1982;
Colbert et al. 1994) also seen in our continuum data (see Figure
1). MERLIN and the Very Long Baseline Array (VLBA) observations of
the two nuclear continuum sources show brightness temperatures of
7 x 10$^6$ K for the northern component and 1.8 x 10$^7$ K for the
southern component (Gallimore \& Beswick 2004). The inverted
spectra at low frequency confirm the AGN nature at each of the
nuclei. The loop results most likely from a bubble front swept up
by a superwind emanating mostly from the southern nucleus
(designated as N1 in Figure 1)(Colbert et al. 1994; Ohyama et al
2000). NGC$\,6240$ exhibits HI and OH absorption against the
nuclear continuum (Baan et al. 1985). Recent HI absorption studies
at 0.3 arcsecond resolution with MERLIN distinguish the absorption
at each of the two nuclear components (Beswick et al. 2001).

ASCA, XMM and Chandra data confirm the presence of two deeply
buried AGNs in the NGC\,6240 system on the basis of a hard X-ray
component with neutral Fe K$\alpha$ lines in addition to the soft
X-ray components due to the starburst (Iwasawa \& Comastri 1998;
Boller et al. 2003; Komossa et al. 2003). The most prominent AGN
is located at the southern nucleus N1, where the obscuration is
the highest. The extended X-ray emission has a close correlation
with the well-known (butterfly-shaped) H$\alpha$ emission.
H$\alpha$ studies with HST confirm the presence of filamentary
structures filling the inner volume of the arc and confirm the
presence of confining walls of the outflow at either side of the
nucleus (Gerssen et al. 2004). Significant H$\alpha$ structures
have also been found in NGC$\,6240$ in the form of a
butterfly-shaped structure that partially superposes the radio arc
and extended radio structure.

NGC$\,6240$ displays strong H$_2$ {\it v} = 1-0 S(1) and [Fe II]
line emission that peaks between the stellar light of the nuclei
but lies closer to the southern nucleus (van der Werf et al. 1993;
Joseph \& Wright 1985; Ohyama et al. 2000). Spectroscopic studies
of H$_2$ at K band allow the separation of the dynamics of the two
nuclei (Tecza et al. 2000). The NIR light of each of the nuclei is
dominated by red super-giants formed during a short episode of
intense star-formation 15-25 million years ago. K band infrared
imaging with Keck II and NICMOS on the Hubble Space Telescope has
revealed elongated structures at both the North and South nuclei
and considerable substructure within each nucleus (Max et al.
2005). Additional point-like regions are found around the two
nuclei, which are thought to be young super-star-clusters.

CO(2$-$1) emission studies with the IRAM interferometer show a
similar structure as the H$_2$ emission and also peaks between the
two nuclei (Tacconi et al 1999). Most of the CO flux is
concentrated in a thick and turbulent disk-like structure between
the two IR/radio nuclei. Studies with Nobeyama Rainbow
interferometer (Hagiwara 1998) indicate that the HCN(1-0) and
HCO$^+$ fluxes also peak between the nuclei, and do not coincide
with the star-formimg region in the galaxy (Nakanishi et al.
2005). The molecular structure accounts for a large fraction
(30-70\%) of the dynamic mass. Although the central location of
the molecular material in NGC$\,6240$ is not unique, it is notably
different from the (more advanced) interaction in Arp 220 where
the emission peaks at the two nuclei (Scoville et al 1997;
Sakamoto et al 1999).

This paper presents studies of the OH and HI absorption in the
NGC\,6240 system using the NRAO Very Large Array (VLA) in
A-Configuration. With NGC\,6240 at a distance of 104 Mpc the
spatial conversion for the VLA data is 504 pc per arcsecond, which
complements the resolution of other spectral line and continuum
studies. Our data reveal more of the dynamics of the system and
provide connections to studies of other atomic and molecular
emissions.

%
%

\section{Observations}

Observations of the HI $21\,$cm line and the OH main lines at
$1665$ and $1667\,$MHz toward NGC$\,6240$ were made on September
1, 1995 with the NRAO Very Large Array in the A-configuration.
Phase tracking was centered on $\alpha(1950) = 16^{h} 50^{m}
28\fs3$, $\delta(1950) = 02^{\circ} 28^{\prime}
53^{\prime\prime}$.

The HI line was observed using a $2\,$IF mode, each IF having a
bandwidth of $6.25\,$MHz subdivided into $32$ channels of
$195.3\,$kHz in width. This setup resulted in a usable velocity
coverage of $1298\,$km$\,$s$^{-1}$ and a velocity resolution of
$43.3\,$km$\,$s$^{-1}$. The spatial resolution in the HI spectral
data is 1.95" x 1.79" for NA weighting and the channel width is
43.3 \kmss. The rms in the individual channel maps is 0.49
mJy\,beam$^{-1}$.

Both OH lines at 1665 and 1667 MHz were observed simultaneously
using two partially overlapping IFs of width 6.25\,MHz with 32
channels of width 195.3\,kHz. The synthesized band used for this
discussion has a center frequency of 1666.38\,MHz (between those
of the 1665 and 1667\,MHz lines) and has a total velocity coverage
of 1326\,km$\,$s$^{-1}$. The velocity scale in this presentation
is heliocentric in the optical definition and has been re-gridded
using the rest frequency of the 1667\,MHz line.  The resolution in
the OH maps is 1.11 x 1.08" and the channel width is 36.7 \kmss.
The rms in the channel maps is 0.41 mJy\,beam$^{-1}$.

Continuum maps at 21 and 18-cm have been constructed using
line-free channels. The rms and the beam size of the two maps are
respectively 0.46 mJy\,beam$^{-1}$ with 1.965" x 1.79" and 0.28
mJy\,beam$^{-1}$ with 1.11" x 1.08".

The data were reduced using the NRAO software package AIPS. The
flux and bandpass calibrator and phase calibrator were 3C 286 and
1648+015 for both HI and OH line data sets. Image cubes were made
with a pixel size of $0\farcs3$, using a variety of weighting
schemes. Continuum data sets were constructed by averaging
line-free channels. Subtraction of the continuum was done
independently in the visibility and in the image domain, resulting
in consistent results. In the case of the OH data both IFs were
imaged independently and joined after the continuum subtraction,
averaging overlapping channels. Due to the uncertainty of the
baseline structure at the edges of the spectrum, we estimate a
flux uncertainty of 20\% for the OH absorption data and 10\% for
the HI absorption and the continuum data.

%
%

\section{Results}

The distance of NGC\,6240 is assumed to be $D$ = 104 Mpc for a
systemic optical velocity of 7275 \kms using $H_0$ = 70 kms$^{-1}$
Mpc$^{-1}$. At this distance the spatial conversion is 504 pc per
arcsecond. In the discussions below, we have adopted the radio
designations of Colbert et al. (1994) for the nuclei N1 and N2 and
the components of NGC\,6240. In addition, the suggestion has been
made for the northern component N3, which may be a third nucleus
or an enhanced fragment of the northern galaxy. There is a
Southern extension S and various West components W1 - W4 forming
the arc structure of 5.9 kpc in size. A W0 component has been
designated in the continuum structure to the west of N3. These
designations have been indicated in Figure \ref{continuum}b.

\subsection{Continuum studies}

\begin{figure}[h!]
\centerline{\includegraphics[angle=0,width=7.5cm]{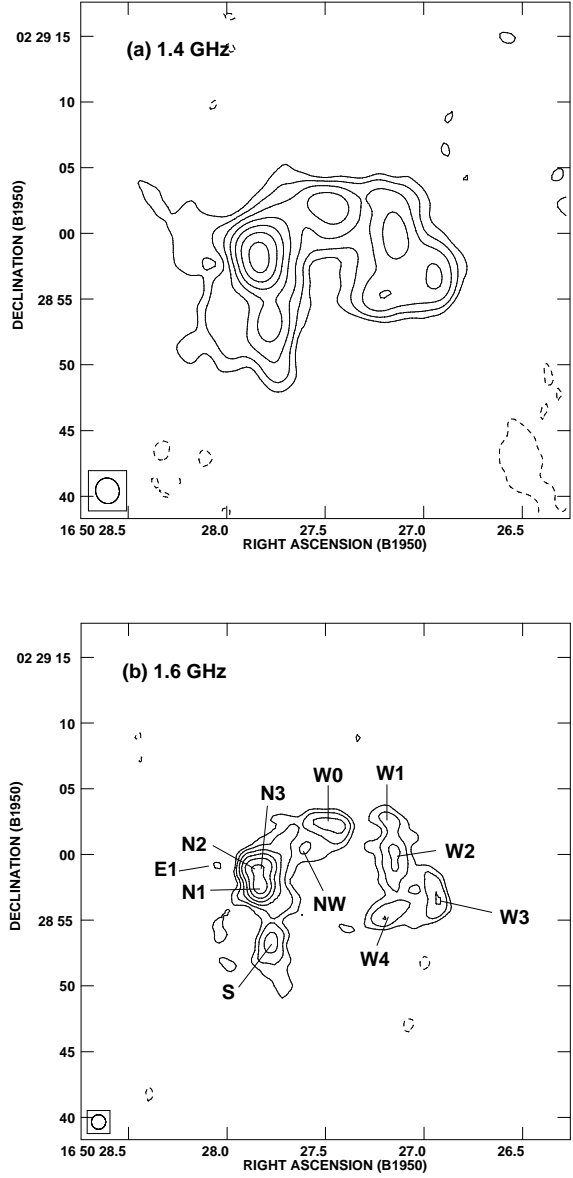}}
\caption{Continuum structure at L-band towards NGC\,6240. {\it
Upper panel}: the 1420\,MHz continuum emission with contour levels
of -1 and 1 to 128 by factors of two times 1.2 mJy\,beam$^{-1}$.
The peak in the map is 108.8 mJy\,beam$^{-1}$. {\it Lower panel}:
the 1666\,MHz continuum emission with contour levels of -1 and 1
to 64 by factors of two times 1.1 mJy\,beam$^{-1}$. The peak in
the map is 55.7 mJy\,beam$^{-1}$. Labelling of components
according to the nomenclature of Colbert et al. (1994).}
\label{continuum}
\end{figure}

Contour maps of the natural-weighted continuum emission at
$1420\,$MHz and the robust-weighted continuum emission at
$1666\,$MHz are presented in Figure \ref{continuum}.  The
integrated flux densities in the maps are 466 and 333 mJy with
peak values 108 and 58 mJy, respectively. Our A-array L-band
images are consistent with previous B-array images except that our
peak fluxes are higher by 10$-$20\%, when convolved to a
resolution of 4.79" x 4.39" of Colbert et al. (1994), possibly due
to slightly different integration boxes. The 21 cm map shown in
Figure \ref{continuum}a (upper panel) is optimized for the
detection of extended low-brightness features and shows that the
individual radio components N1 (South) and N2 (+N3) (North) are
embedded in a halo of diffuse emission. The higher resolution 18
cm map in Figure \ref{continuum}b (lower panel) clearly separates
the two nuclei N1 and N2 (+N3). At higher resolution the N1$-$N2
axis is found to be at PA = 20$^o$ with a (projected) separation
of the nuclei of 1.575" or 93 pc (Hagiwara et al. 2003; Gallimore
\& Beswick 2004).

The structure along the western arm (W1-W4) shows diffuse emission
without any sharp peaks at this resolution. The new components NW
and W0 have been added to indicate the diffuse structure west of
N3. Diffuse extensions can be seen in the north-east NE, the
south-east SE, and an S extension. We present radio continuum
parameters derived from the $1666\,$MHz map in Table~1.

The large-scale continuum structure shows a strong similarity to
the butterfly structure found in the optical and X-ray (Komossa et
al. 2003). The primary cause of this structure would be a nuclear
blowout from the nuclear region of N1 in the southern galaxy. The
large-scale H$\alpha$ structure towards the west (see Max et al.
2005) appears to match and complement the loop structure in the
radio. In addition, there eastern complement to the western loop
in H$\alpha$ and soft X-ray (Max et al. 2005; Komossa et al.
2003). The SE and NE radio extensions agree with X-ray and
H$\alpha$ structure and form "the base" of a similar blowout
bubble to the east of the nuclei. The S and NW radio extensions in
Figure \ref{continuum}a,b have counterparts also in the larger
scale H$\alpha$, H$_2$ and X-ray emission structures.
\subsection{The HI Absorption}
\begin{figure}[!ht]
\includegraphics[angle=0,width=16.5cm]{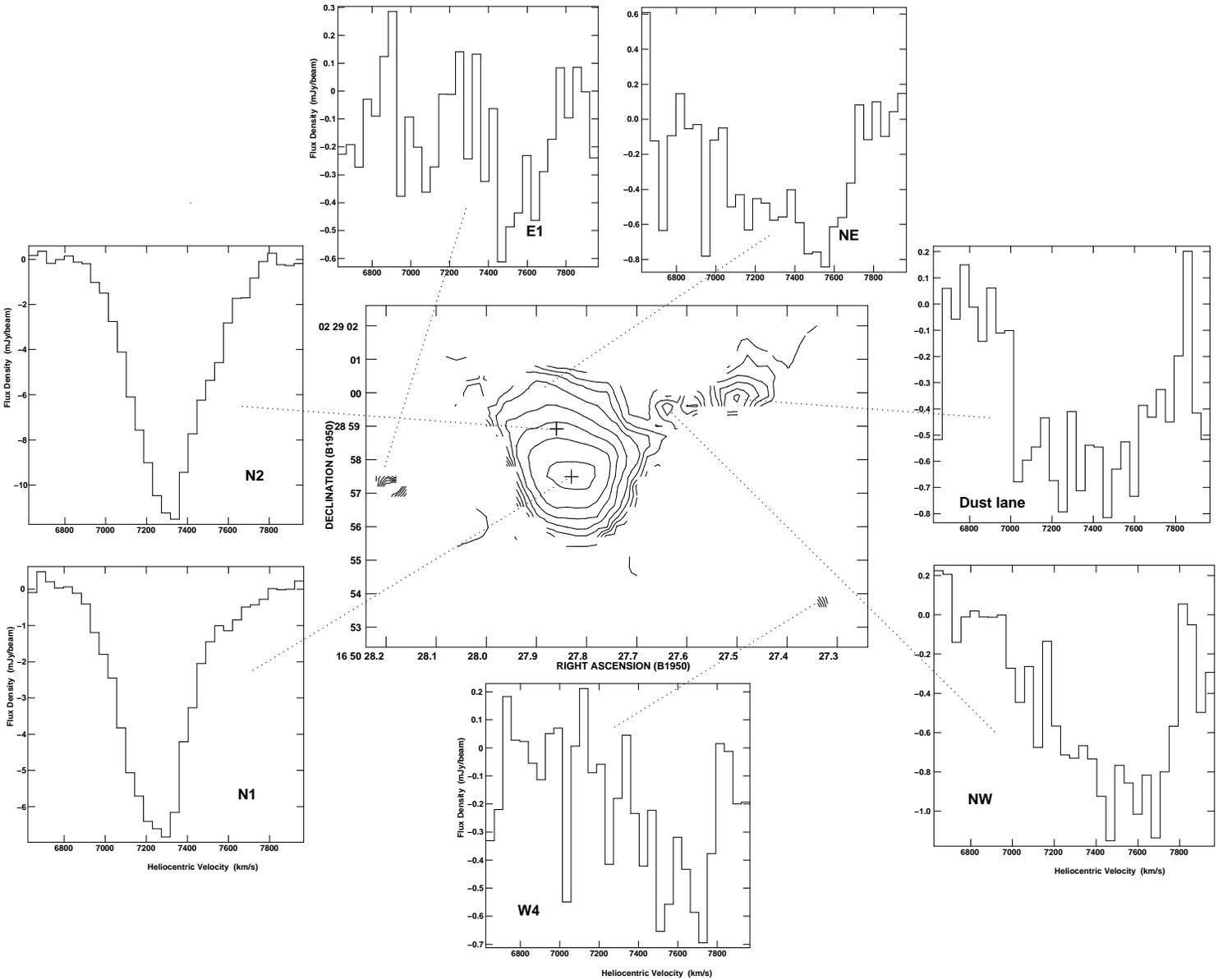}
\caption{Spectral signatures of the HI absorption in NGC$\,$6240.
The zeroth moment map of integrated HI absorption is presented in
the central frame and is similar to Figure \ref{hicolumn}. The
spectra at seven locations have velocity tick marks of 6800 to
7800 \kmss. We note that the significance of the spectral profiles
is low at some locations in the column density maps.}
\label{hispec}
\end{figure}

{\bf The HI line characteristics} - The HI absorption spectra in
the extended emission region of NGC\,6240 have been given in
Figure \ref{hispec}. The profiles of the HI absorption at the two
nuclei are very similar as both spectra have a FWZI width of about
900 \kms and a half-power width of about 348 \kmss. However, the
absorption at N2 is 1.7 times stronger that the absorption at N1,
and the line at N2 is more symmetric than the N1 line that is
skewed due to a higher velocity component. The systemic velocities
at N1 and N2 are 7295 \kms and 7339 \kms respectively. N1 lies
just south of the peak in the HI absorption column density in
Figure \ref{hispec}.

The spectra presented in Figure\ref{hispec} indicate that there
are large differences in the absorbtion columns and that
absorption is seen over a velocity range of more than 900 \kmss.
This velocity width results from the rotation in each of the
galaxies, the orbital velocity component of the two galaxies, and
the inflow and outflow due to the interaction. The line of sight
to each nucleus does not provide an accurate estimate of the
systemic velocity of that nucleus. The spatial resolution of the
21-cm data is 980 x 900 pc and our line-of-sight towards the two
nuclei will sample multiple velocity components. The
high-resolution HI absorption study with MERLIN at 0.3 arcsec
resolution by Beswick et al. (2001) revealed two isolated
absorption components at the locations of the N1 and N2 nuclei at
velocities 7087 and 7260 \kms (radio definition). Using the
optical heliocentric definition we find systemic velocities V(N1)
= 7258 \kms and V(N2) = 7440 \kms and we adopt these as a more
accurate approximation of the systemic velocities of the two
nuclei. It should be noted that these velocities straddle the
absorption peak in the two nuclear absorption spectra of Figure
\ref{hispec}. While the peak of the absorption column density
coincides with N1, the centroid absorption velocity at N1 is about
100 \kms lower, due to the lower velocity of the structural
component between the nuclei.

\begin{figure}[!ht]
\begin{center}
\includegraphics[angle=0,width=7cm]{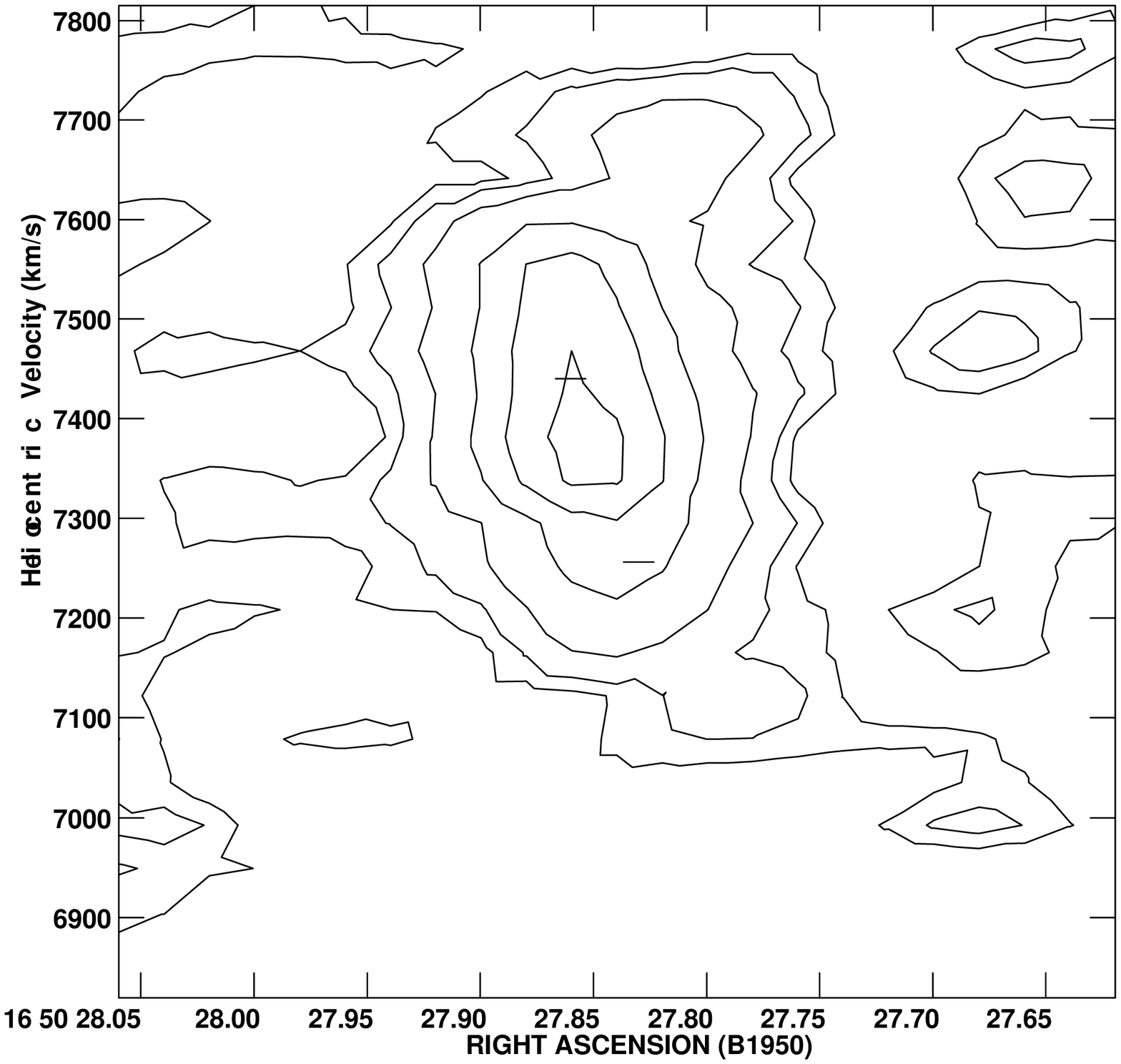}\\[5mm]
\includegraphics[angle=0,width=7cm]{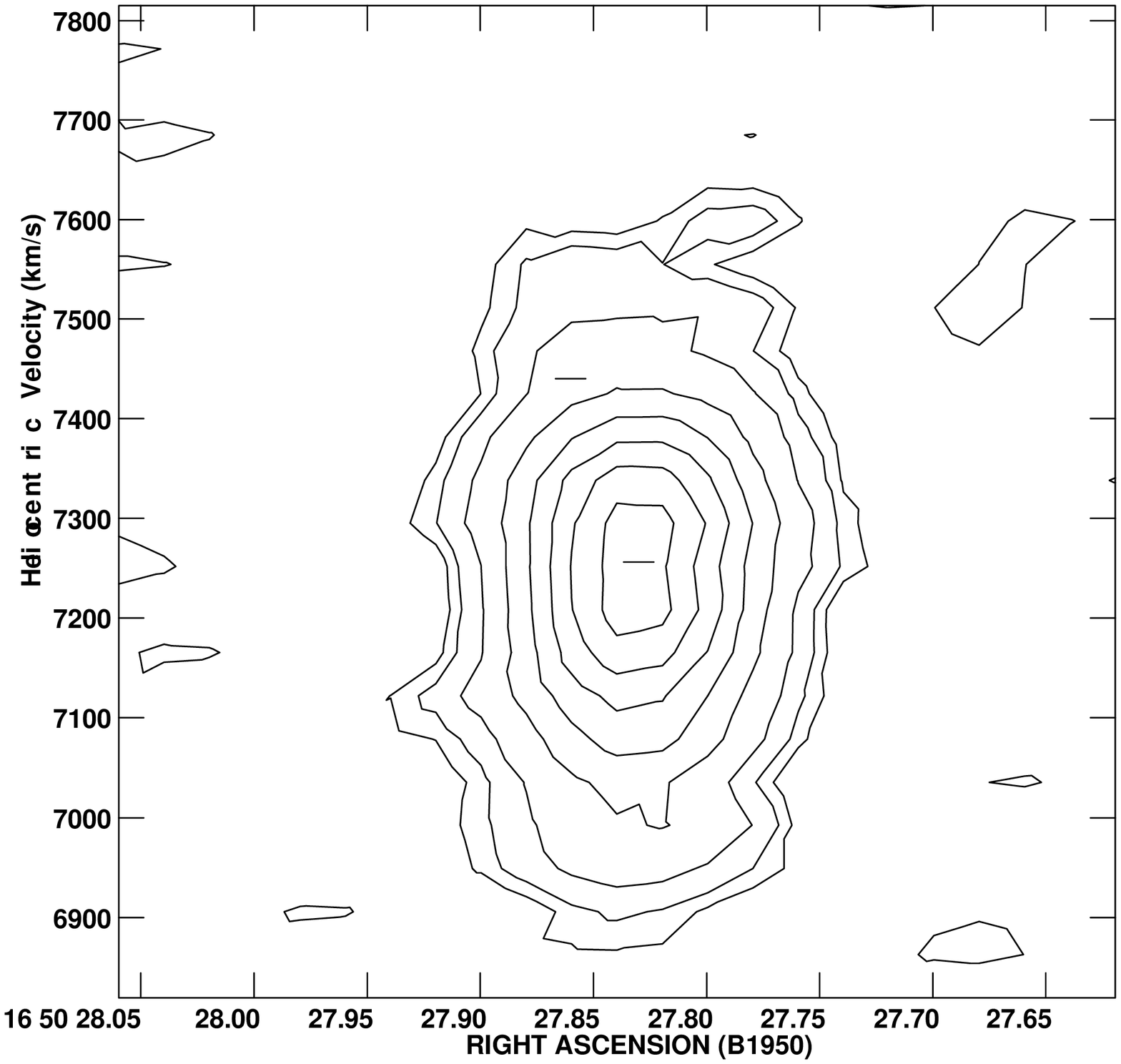}
\end{center}
\caption{The HI velocity-position diagrams in the nuclear region
for two declinations. The systemic velocities of N1 and N2 are
7258 and 7440 \kmss. {\it Upper diagram:} Close to nucleus N2 at
declination 02 28 59.9. {\it Lower diagram:} Close to nucleus N1
at declination 02 28 57.2. For an rms of 0.45 mJy\,beam$^{-1}$,
the contour levels are at 0.5, 1, 2, 4, 6, 8, 10, 12, 14, 16
mJy\,beam$^{-1}$. The declination-velocity locations of the two
nuclei have been marked in the diagrams. Features at the lowest
contours have a marginal significance. } \label{hipvd}
\end{figure}

\begin{figure}[htb]
\begin{center}
\includegraphics[angle=0,width=5.2cm]{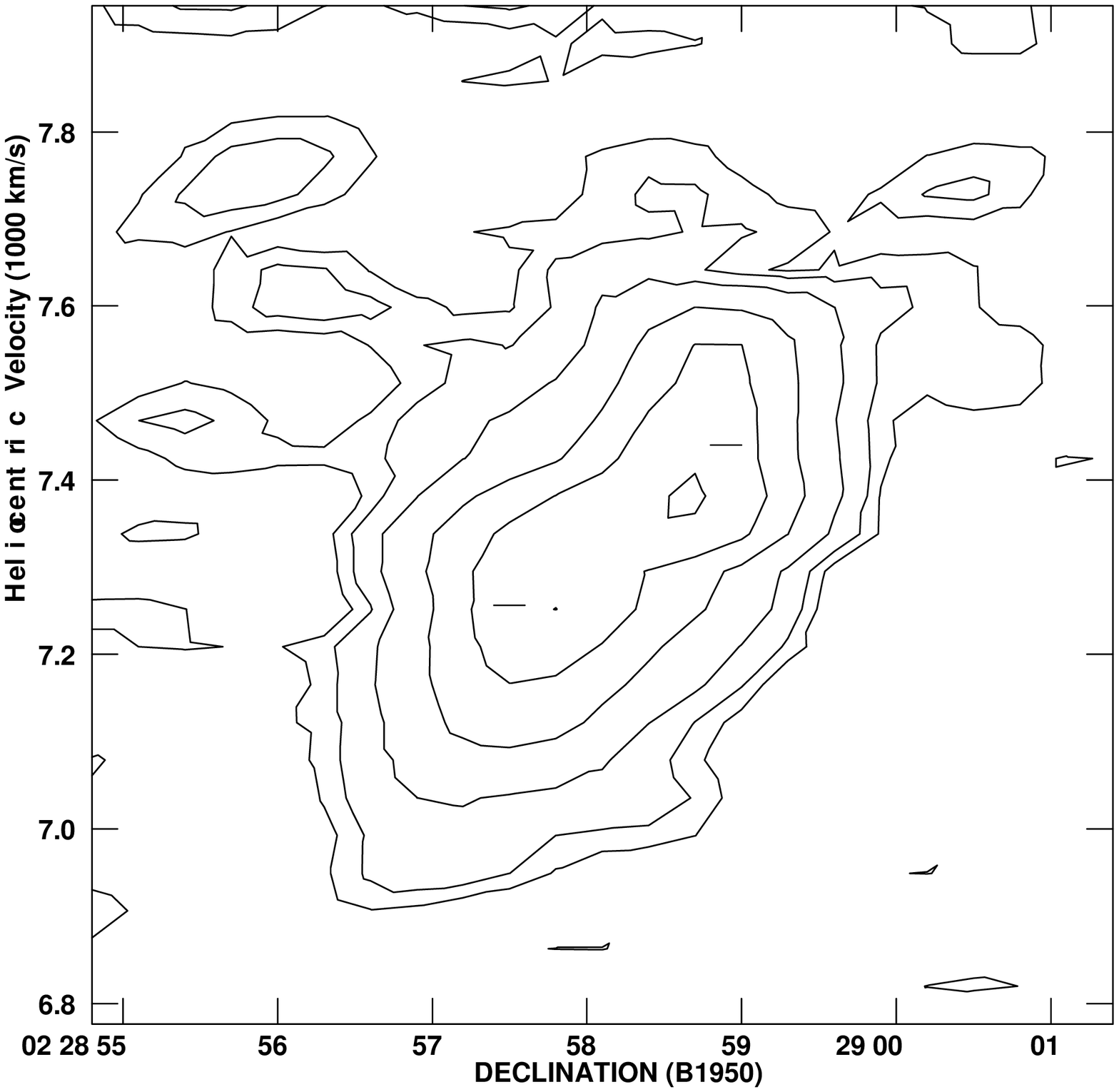}
\includegraphics[angle=0,width=5.2cm]{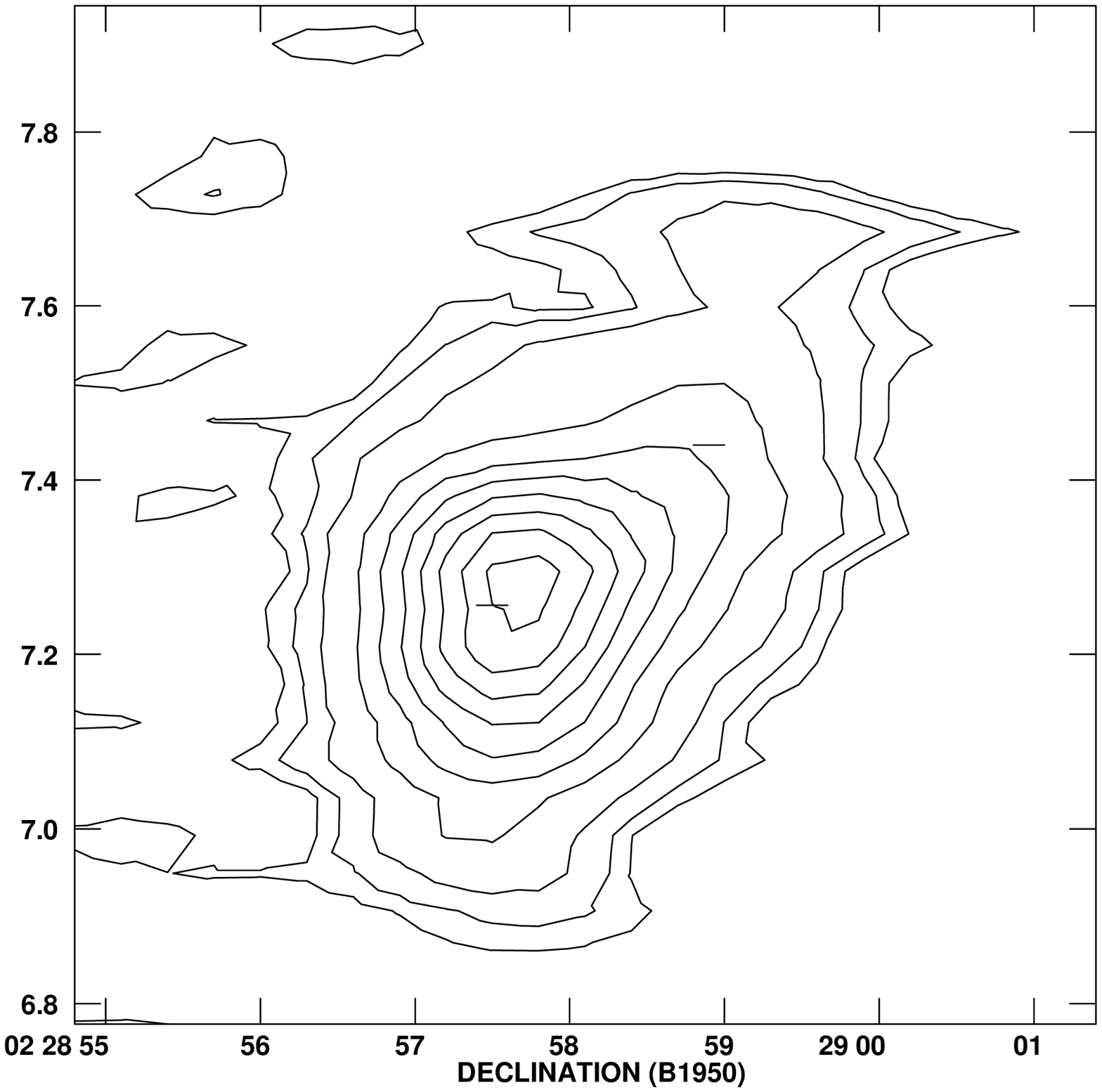}
\includegraphics[angle=0,width=5.2cm]{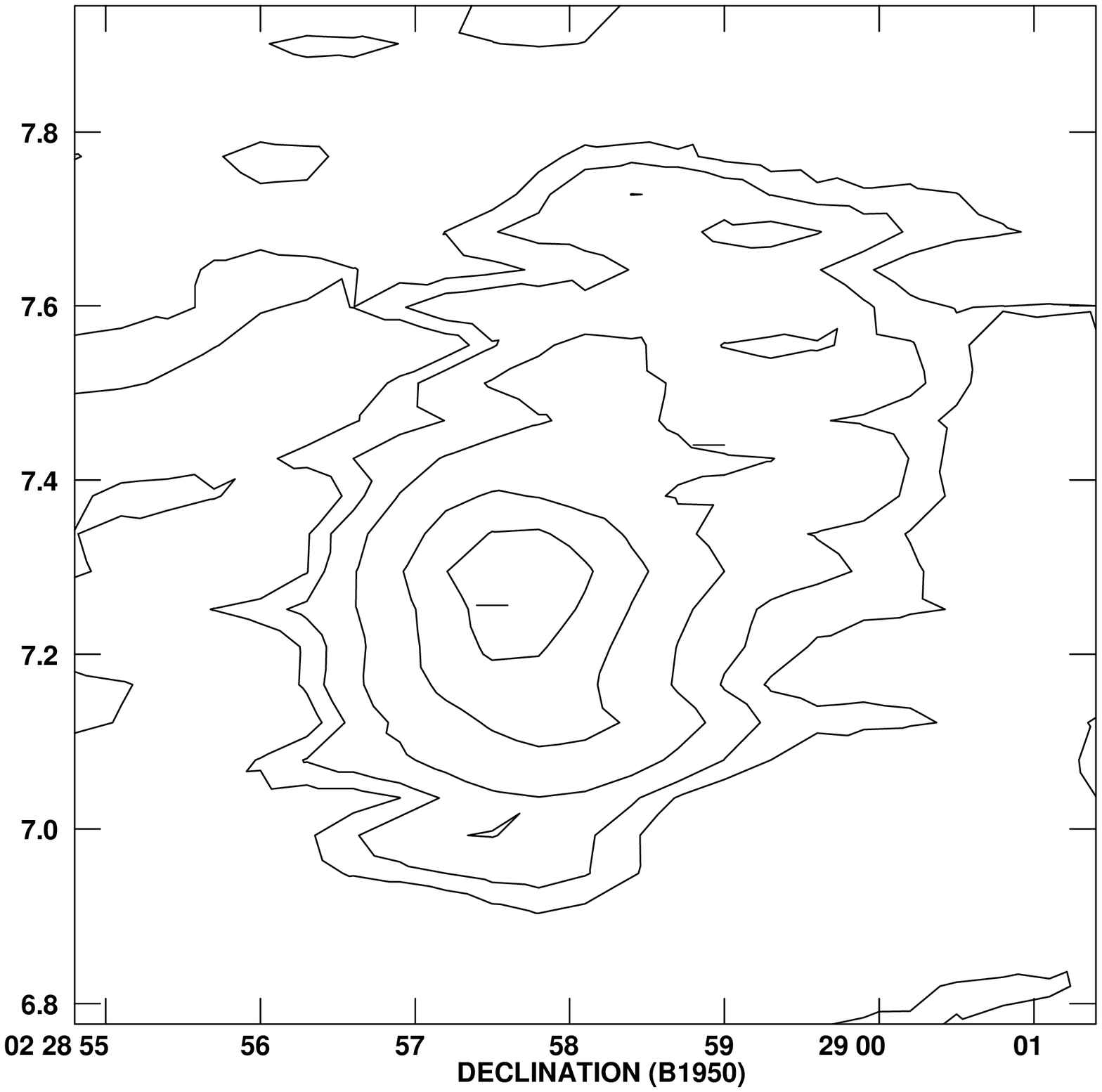}
\end{center}
\caption{The HI velocity-position diagrams in the nuclear region
for three RA values. The systemic velocities of N1 and N2 are 7258
and 7440 \kmss. {\it Left diagram:} Just east of nucleus N2 at RA
= 16 50 27.88. {\it Middle diagram:} Just west of nucleus N1 at RA
= 16 50 27.82. {\it Right diagram:} West of nucleus N1 and N2 at
RA = 16 50 27.78. For an rms in these maps of 0.45
mJy\,beam$^{-1}$, the contour levels are at 0.5, 1, 2, 4, 6, 8,
10, 12, 14, 16 mJy\,beam$^{-1}$. The declination-velocity
locations of the two nuclei have been marked in the diagrams.
Features at the lowest contours have a marginal significance. }
\label{hipvr}
\end{figure}

{\bf The HI PV diagrams} - Figures \ref{hipvd} and \ref{hipvr}
present velocity-position maps in two principal E-W and N-S
directions. These diagrams cover only the double-nucleus region.
The RA$-$velocity diagrams of Figure \ref{hipvd} show the velocity
structure in the south and the north of the system. Close to N1
where the column density peaks at 7265 \kmss, the rotation is
essentially south-to-north with a gradient to be determined from
the PV plot along the declination axis. In the northern PV
diagram, the extended absorption peaks just south of N2  at 7385
\kms and shows an west-to-east velocity gradient of 1.0 \kmss
pc$^{-1}$ close to N2. In the North, an additional (weak) second
component appears at 7700 \kms with a east-to-west velocity
gradient of 0.53 \kmss pc$^{-1}$. The extended low-declination
outflow structure south of N1 reaches 6900 \kms (Fig.
\ref{hipvd}b) and has a southwest-to-northeast gradient of 0.89
\kmss pc$^{-1}$.

The declination$-$velocity diagrams of Figure \ref{hipvr} display
three velocity profiles along the east side, close to the center,
and the west side of the central region (along a south-north RA
direction). The diagrams show a changing south-to-north velocity
gradient resulting from three distinct components. South of N1
there is a gradient of 1.49 \kmss pc$^{-1}$; north of N2, a
gradient of 1.15 \kmss pc$^{-1}$. In between N1 and N2 the
gradient is 0.26 \kmss pc$^{-1}$, which is close to the predicted
value of 0.24 \kmss pc$^{-1}$ based on the velocity difference of
the two nuclei. This central component is dominated by an
accumulation of gas in between the two nuclei. A similar
absorption structure is found in the OH data. In accordance with
Fig. \ref{hipvd}, an east-to-west component enters between 7300
and 7750 \kmss, which occurs at high declination at the west side
(right frame) of the source, i.e. going towards the NW region and
possibly representing streaming gas motions in the northern
galaxy. A number of rather marginal but distinctly offset
components are found west of N2 (Fig. \ref{hipvr}a) and southeast
of N2 (Fig. \ref{hipvd}a), covering a large velocity range of
7000$-$7750 \kmss.

In our discussion of the OH absorption data below, we will
correlate our findings of the HI absorption in the nuclear N1$-$N2
region. However, the OH absorption is solely confined to the
nuclear region and does not display any of the extensions found in
this section.

The combined PV diagrams show the existence of five independent HI
components found against the nuclear region: (1) a disk-like
structure with three components with distinct gradients of 1.49
\kmss pc$^{-1}$ south of N1, of 1.15 \kmss pc$^{-1}$ north of N2,
and of 0.26 \kmss pc$^{-1}$ between N1 and N2. A single gradient
covering the whole range would have a south-to-north gradient of
0.32 \kmss pc$^{-1}$.; (2) the region north of N2 shows a second
(reverse) southwest-to-northeast gradient of 1.0 \kmss pc$^{-1}$;
(3) a high-declination east-to-west structure between 7250 and
7750 \kms with a gradient of 0.53 \kmss pc$^{-1}$ providing a
connection with the NW absorption region; (4) an outflow component
reaching 6900 \kms associated with nucleus N1 with a
southwest-to-northeast gradient of 0.89 \kmss pc$^{-1}$; and (5)
some distinct (but marginal) offset components covering
7000$-$7750 \kms associated with the disturbed region
west-south-west of the nucleus N1.

\begin{figure}[!h]
\begin{center}
\includegraphics[angle=0,width=8cm]{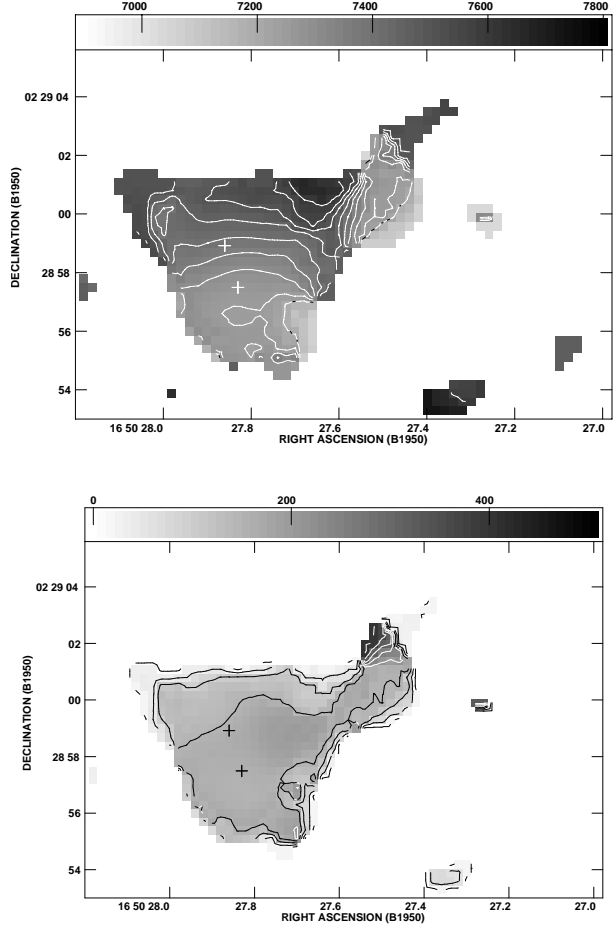}\\
\includegraphics[angle=-90,width=8cm]{f5b.eps}
\end{center}
\caption{HI velocity moment maps. Positions of N1 and N2 are
marked by crosses.  {\it Upper diagram}: HI 1st moment map.
Contours are plotted at 7200, 7250, 7300, 7350, 7400, 7450, 7500,
7550, and 7600 \kmss. The grey-sale is from 6900 to 7800 \kmss.
{\it Lower diagram}: HI 2nd moment map, showing HI velocity
dispersion. Contours are plotted at 10, 50, 100 to 500 by 50
\kmss. The grey-scale starts at 0 \kms and ends at 500 \kmss. The
central region has velocity half-widths greater than 150 \kmss.}
\label{himom12}
\end{figure}

{\bf The moment maps} - The first moment HI map in Figure
\ref{himom12}a confirms the dominant south-to-north (component
(1)) velocity gradient along the N1 - N2 axis suggesting organized
rotation. The velocity gradient deduced between N1 and N2 is 0.18
\kmss pc$^{-1}$, which is smaller than the one obtained above from
the PV diagrams.

A large-scale east-to-west rotation (component (3)) in the
northern region causes curved iso-velocity lines and continues
into the NW region but is interrupted by a lower-velocity
north-south component crossing the region at W0. This interruption
has the signature of the foreground dust lane passing just west of
the nuclei in optical images. The dust lane absorption at W0 is at
7400 \kmss, which is close to the systemic velocity of N2.

At the locations of the two nuclei, the systemic velocities in
Figure \ref{himom12}a are V(N1) = 7235 \kms and V(N2) = 7305
\kmss. The velocity difference of 70 \kms is smaller than the 120
\kms found in the PV diagrams of Fig. \ref{hipvd}, which would
come closer to the difference of 182 \kms found at higher
resolution (Beswick et al. 2002) and in other molecular data.
Outside the main structure we can also identify absorption at W2
at V = 7034 \kms and at the eastern edge of W4 at 7687 \kmss,
which needs confirmation.

The second moment map in Figure \ref{himom12}b shows a rather
curious structure with a band of large velocity widths (more than
150 \kmss) running from N1 and N2 into the NW region. For
comparison, the MERLIN line widths are largest at N2 with 300 km/s
and narrowest towards the northwest with 60 \kms (Beswick et al.
2001).

\begin{figure}[!h]
\begin{center}
\includegraphics[angle=0,width=8cm]{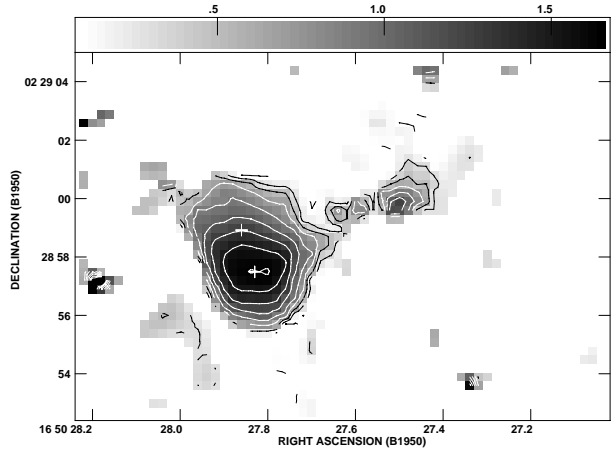}
\end{center}
\caption{HI absorption column density map. Contours for N$_{\rm
H}$ are 0.1, 0.3, 0.5, 0.7, 0.9, 1.1, 1.3, 1.5, and 1.64 and grey
scale from 0.1 to 1.64 in units of 7.78 $\times$ 10$^{21}$
cm$^{-2}$ assuming T$_S$ = 100 K.} \label{hicolumn}
\end{figure}

A distinct low-velocity component (seen also as 4 and 5 in the PV
data) located at west-south-west of N1 is likely related to the
superwind outflow. This feature at PA = 25 $^o$ appears to have an
west-to-east velocity gradient and lower line widths. It should be
noted that this structure also appears at the same location in the
OH data. Furthermore, the position angle of the outflow component
also points to the eastern edge of the W4 component that is also
found in the moment maps of Fig. \ref{himom12} and \ref{hicolumn}.
In addition, the offset components found at the outflow position
in the PV diagrams have a velocity range of 7000$-$7750 \kmss,
indicating a highly disturbed region.

{\bf The HI absorption column density} - The absorption column
density presented in Figure \ref{hicolumn} and in Table 2 is
determined using the absorption line strengths and the associated
continuum data at 1420 MHz (Figure \ref{continuum}a). The
expression for the hydrogen column density used is $N_{\rm
H}$/$T_{\rm S}$ = 1.823  $\times$ 10$^{18}$ $\int$ $\tau(V)$ $dV$
cm$^{-2}$, where $T_{\rm S}$ is the hydrogen spin temperature, and
$\tau$ is the absorption optical depth. The map shows the highest
column density of $N_H$ = 1.28 x 10$^{22}$ cm$^2$ at nucleus N1
using a spin temperature of 100 K. The optical depth is largest at
N1 with 0.15 ($\pm$ 0.001) and at N2 with 0.11 ($\pm$ 0.001).

The column density map also shows absorption at W0 with 9.42 x
10$^{21}$ cm$^2$, and at W4 with 7.78 x 10$^{21}$ cm$^2$. If one
assumes that the absorption at W0 is composed of a continuation of
the NW absorption plus a contribution of the foreground dust-lane
at a significantly lower velocity, then the dust-lane specific
column density is estimated at 3.2 x 10$^{21}$ cm$^{-2}$. The NW
structure at PA = $-$60$^o$ is located at some 2 kpc from the
nuclei and W0 is at 2.7 kpc. In addition, the NE elongation at PA
= 45$^o$ extends almost 2.0 kpc from N2. The absorption spot W4 is
4.2 kpc away from N1.

\subsection{The OH Lines}

\begin{figure}[!h]
\centerline{\includegraphics[angle=-90,width=7.5cm]{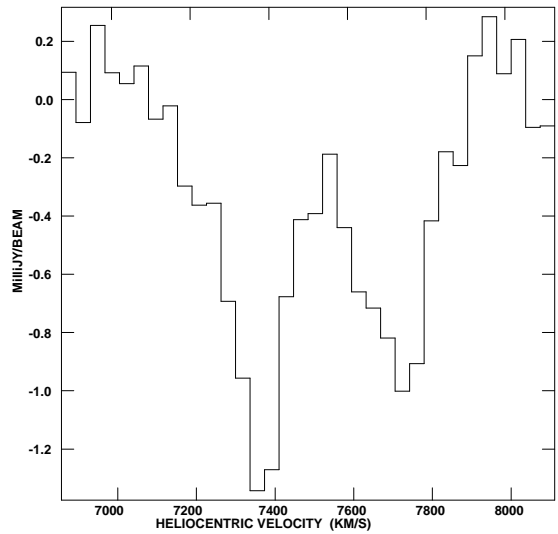}}
\caption{Integrated spectrum of OH absorption taken across both
nuclei with the 1667 MHz (left) and the 1665 MHz (right at +351
\kmss). The velocity axis of the spectrum is in the rest frame of
the 1667 MHz line.} \label{ohspec}
\end{figure}

{\bf The OH line characteristics} -  The integrated spectrum of
the 1667 MHz and 1665 MHz OH absorption is depicted in Figure
\ref{ohspec} in the rest frame of the 1667 MHz line. The adopted
systemic velocities of the two nuclei of 7258 and 7440 \kms lie
just below and above the absorption peaks of the two lines. The
integrated OH spectrum across the whole source shows an overall
line ratio of 1.3 for the two absorbing transitions. Channel maps
of the OH data cube have not been displayed because the spectral
information is better presented by other means.

The shallowness and irregularity of the single-dish OH absorption
spectrum (Baan et al. 1985) suggested that some of the absorption
in the system had been filled in with emission. With this in mind,
the OH data have been scrutinized in a search for OH emission but
no clear evidence has been found. The two absorption lines in Fig.
\ref{ohspec} suggest an asymmetry of the line profiles and
non-similarity that could indeed be explained by a partial
infilling with emission on the high velocity side of the 1665 MHz
line and/or at the low side of the 1667 MHz line.

\begin{figure}[!h]
\centerline{\includegraphics[angle=0,width=7.5cm]{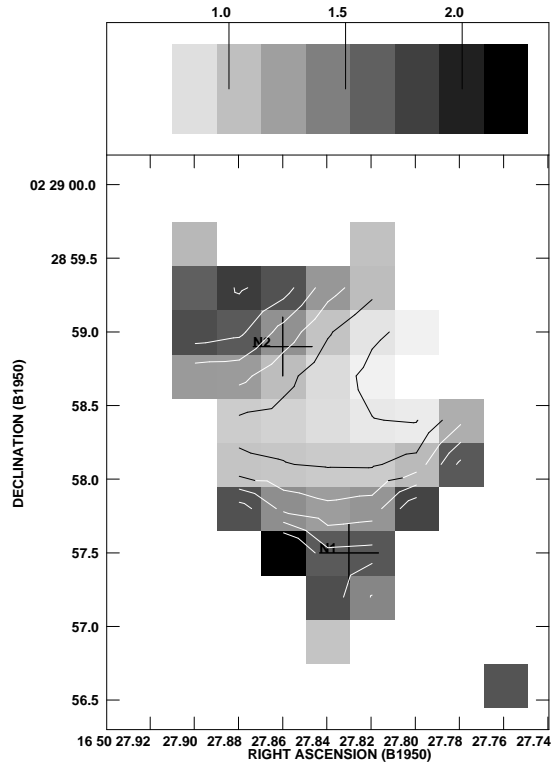}}
\caption{The hyperfine line ratio of the 1667 and 1665 MHz
absorption lines across the nuclear region.  The location of the
two nuclei are indicated in the diagram. The contour levels are
0.8$-$1.8 with intervals of 0.2. The lowest contour of 0.8 is at
the west side of the central region and subsequent contours are
increasingly further out towards the two nuclei. Values of 1.6 are
found at N1 and 1.2 at N2. The highest optically$-$thin ratio of
1.8 is found north of N1, while south of N1 the ratio decreases
again to 1.4. } \label{ohrat}
\end{figure}

The OH hyperfine ratio in the nuclear absorption region is
depicted in Figure \ref{ohrat} using the sums of the channel maps
for the 1667 and 1665 MHz lines as depicted in Figure
\ref{ohspec}.  The central region between the nuclei exhibits
values below 1.0 and going below 0.8 on the west side. Locations
outside the central region have values in the optically-thick and
optically-thin LTE range of 1.0$-$1.8. The values of the hyperfine
ratios at N1 and N2 are respectively 1.6 and 1.2. The occurrence
of non-LTE conditions in the velocity range of 7330 \kms and below
may be explained with emission in the 1667 MHz line. The range
around 7300 \kms corresponds to a missing (low-velocity) shoulder
of the 1667 MHz line profile (see Fig. \ref{ohspec}).

\begin{figure}[!h]
\centering
\includegraphics[angle=0,width=8cm]{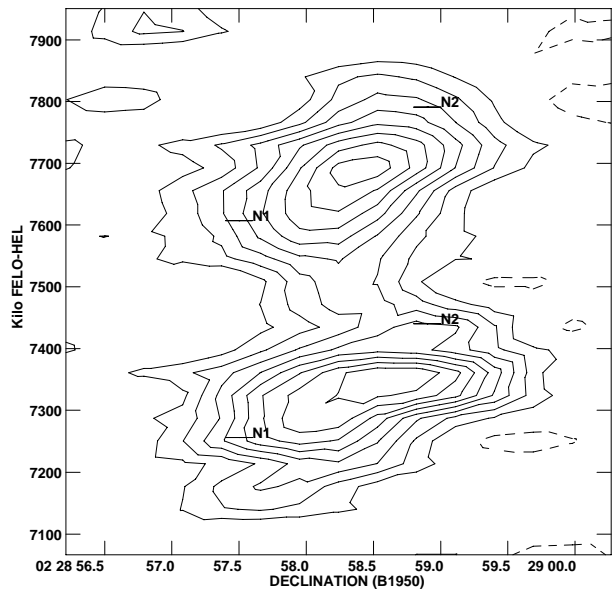}
\caption{Position-velocity map of the 1667/1665 MHz OH absorption
along P.A.=20$\degr$ along the N1$-$N2 line. The spectrum has been
inverted for presentation.  The RA - velocity position of each of
the nuclei has been indicated in the diagram. For an rms of 0.29
mJy\,beam$^{-1}$, the contours are plotted at -1, 1, 2, 3, 4, 5,
6, 7, and 8  $\times$ 0.55 mJy\,beam$^{-1}$. Features at the
lowest contour have marginal significance.} \label{ohpv}
\end{figure}

{\bf The OH PV maps} - The OH data shows substantial absorption
only against the nuclear radio double source and no large
extensions can be found outside the nuclear area. A single OH
velocity-position map is presented at PA = 20$^o$ along the
N1$-$N2 axis (Fig. \ref{ohpv}). Different than that of the HI
absorption, the bulk of the OH absorption occurs between the two
nuclear sources at 7325 \kmss, and shows a dominant south-to-north
velocity gradient in the central region of 0.19 \kms pc$^{-1}$,
which is somewhat smaller than that of the central HI component.
Similar to the HI absorption, there is a changing velocity
gradient across the region with two separate components with
velocity gradients outside N1 and N2. The gradients in the two
shoulders in Fig. \ref{ohpv} south of N1 and north of N2 are
estimated to be 0.75 \kms pc$^{-1}$, which is much steeper than
the central part but still lower than that of the HI estimate.
This component has also a counterpart in the 1665 MHz lines and is
related tot the connecting bridge between the 1665 and 1667 MHz
lines.

In the northern region close to N2, there is a distinct component
with an (estimated) opposite west-to-east velocity gradient of
0.30 \kms pc$^{-1}$. A similar structure has been found in the HI
data (Fig. \ref{hipvr}), which has been associated with rotation
due to the northern galaxy along the N2$-$NW line. In the spectrum
of Figure \ref{ohspec} this translates into the low-velocity
shoulder of the 1667 MHz line.

\begin{figure}[htbp]
\centering
\includegraphics[angle=0,width=7.5cm]{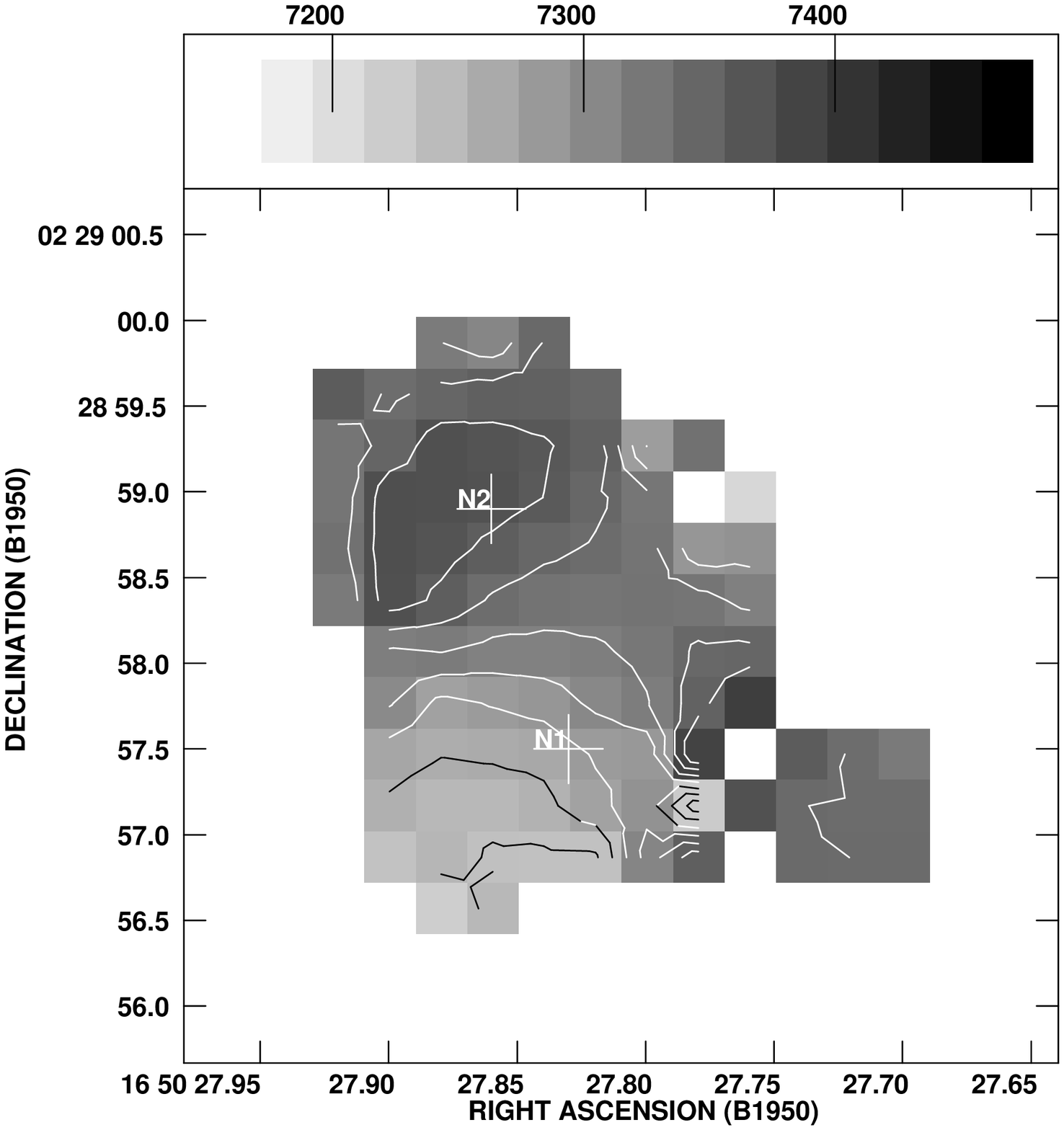}\\[5mm]
\includegraphics[angle=0,width=7.5cm]{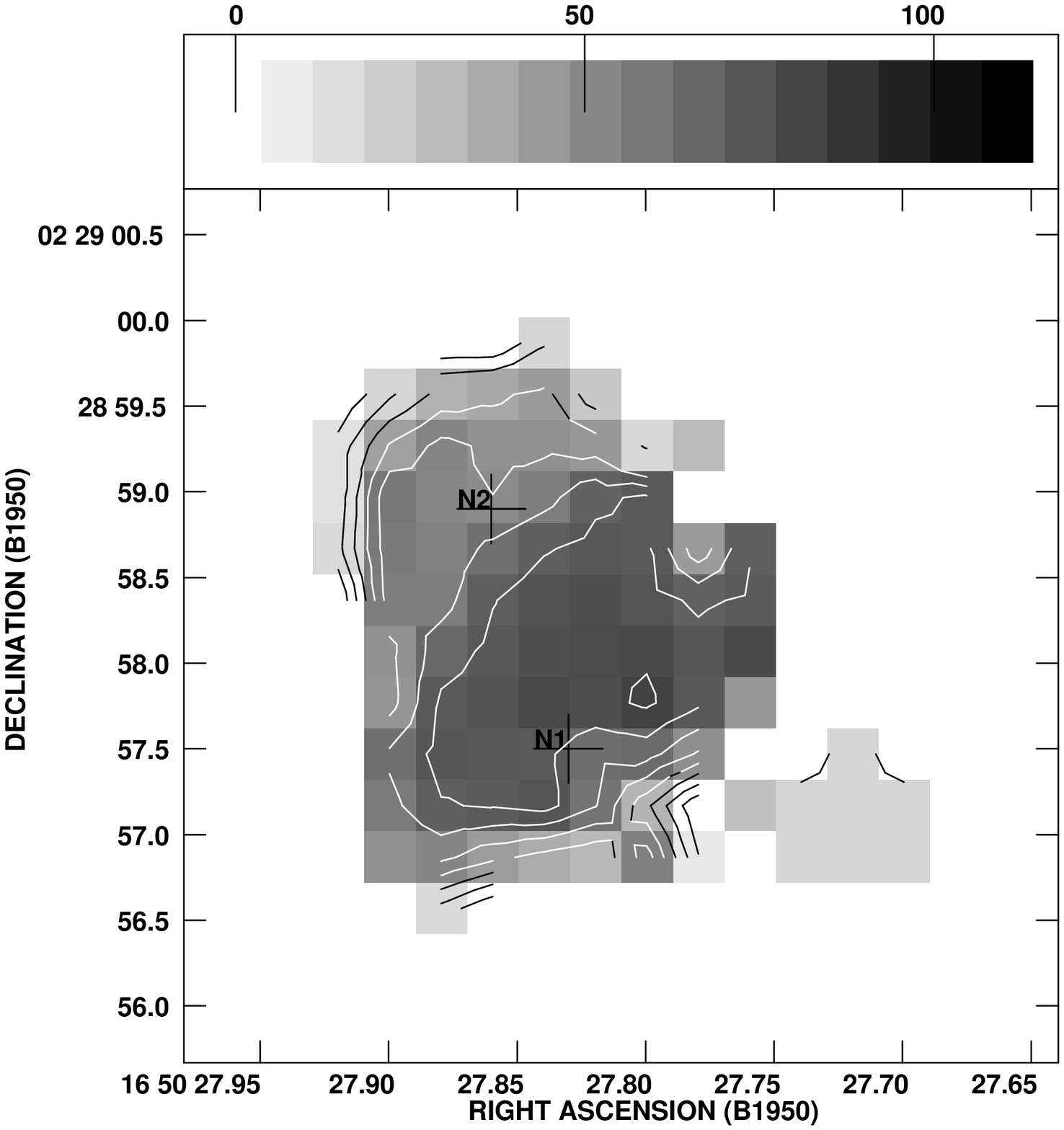}
\caption{First and second moment maps of the OH 1667 MHz line. The
locations of the twin-nuclei are marked by crosses. {\it Upper
diagram} First moment: The line velocity contours superposed on
the grey-scale are spaced linearly by 20 \kms beginning at 7240
\kms and ending at 7360 \kmss. The peak value is 7393 \kms and the
values at N1 and N2 are 7275 and 7370 \kmss. Grey-scale range is
from 7150 \kms to 7400 \kmss. {\it Lower diagram} Second moment:
The line width contours are at 20 to 80 \kms with increments of 10
\kmss. The central region has a line width of 80+ \kmss. }
\label{ohmom12}
\end{figure}

{\bf The moment maps} - The first moment map of the 1667 MHz line
in Figure \ref{ohmom12}a shows a smooth velocity gradient, that
resembles and confirms the HI characteristics in the central
region. The velocity gradient starts in the SE region close to N1
as part of the southern galaxy and continues via N2 into the
northeast direction. There is some evidence of a superposed
east-to-northwest gradient starting at N2 that is associated with
the northern galaxy. The velocities derived for N1 and N2 from the
second moment map are 7255 and 7370 \kmss.

The line width in the 1667 MHz line displayed in Figure
\ref{ohmom12}b is largest at a location between the two nuclei
similar to the HI case, but with a value of 80+ \kms it is
significantly smaller than the 150+ \kms width found in HI. It
should be noted that the highest line widths coincide partially
with the region of non-LTE (super optically-thin with ratio $\leq$
1.0) excitation in Figure \ref{ohrat}. The moment maps of the 1665
MHz lines are all consistent with those of the 1667 MHz.

Figures \ref{ohmom12} also display the curious structure southwest
of N1 at PA = -25$^o$ that is also present in the HI data, and
represents the direction of a jet or is part of a wider nuclear
outflow. At that location the velocity field is confused and the
OH linewidths become narrower. Further to the west there is an
additional (disjoint) region with of very low 20 \kms linewidth at
7360 \kmss, which may relate to the (streaked) extensions at low
declination (57.0") in the PV diagram (Fig. \ref{ohpv}).

\begin{figure}[!h]
\centering
\includegraphics[angle=0,width=7.5cm]{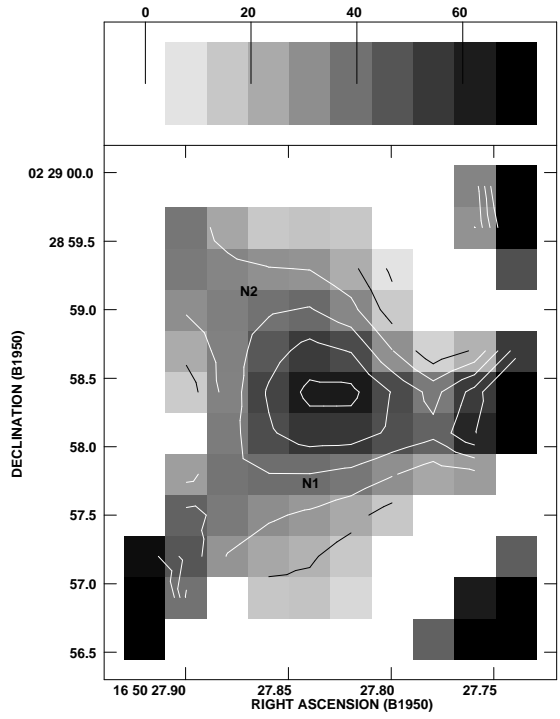}
\caption{OH column density map of 1667 MHz. The integrated column
density contours are 1, 2, 4, 6, 8, 10, and 12 times 8.62 x
10$^{14}$ cm$^{-2}$ using an excitation temperature of $T_{ex}$ =
20 K. The grey scale displays the corresponding optical depth on a
scale of 0 to 0.06. The peak optical depth is 0.063 with a column
density of 1.08 x 10$^{16}$ cm$^{-1}$. } \label{ohcolumn}
\end{figure}

{\bf The OH column density} - The OH column density has been
presented in Figure \ref{ohcolumn} and has been based on the 1667
MHz optical depth using the 18 cm continuum map (Fig.
\ref{continuum}) and the expression N$_{\rm OH67}$ = 2.35 x
10$^{14}$ T$_{ex}$ $\int\tau(V) dV$, where the excitation
temperature T$_{ex}$ has a typical value of 20 K. The region with
the highest OH column density of N$_{\rm OH67}$ = 1.08 x 10$^{16}$
cm$^{-2}$ occurs halfway between N1 and N2. The column densities
at N1 and N2 are a factor of about 1.8 lower. The peak OH optical
depth of 0.063 is a factor of 2 smaller than that of HI.

\section{Discussion}

\subsection{The central gas concentration}

The central gas concentration is clearly present in the OH data,
where it peaks between the nuclei at about 0.9" north of N1 (and
closer to N2), and also in the HI data, where the peak occurs
close to N1. In a projection scenario with N1 being located behind
N2 (see section below), the largest column densities should occur
at N1 and gas distributions of the two galaxies are displaced by
only 790 pc. Therefore, the column differences between the
absorption peaks and the nuclei of 1.3 for HI and 1.8 for OH could
be accommodated by a superposition of two galactic gas
distributions. However, the velocity gradient in the central
region of 0.19 \kmss pc$^{-1}$ for OH and 0.26 \kmss pc$^{-1}$ for
HI is close to the predicted value of 0.24 \kmss pc$^{-1}$, that
results from the velocity difference of the two nuclei. The
central molecular structure is also not tied simply to the orbital
motion of the two nuclei, because the emission peak lies off the
N1-N2 connecting line. The central gas structure could thus be a
superposition of disk gas, which is combined with gas pulled out
of the two galaxies during the interaction and deposited close to
the center of mass of the system. The structure appears
co-rotating within the two nuclear regions.

The centrally peaked OH absorption shows rough agreement with the
findings for other thermally excited molecules such as the CO $J$
= 2$-$1 and H$_2$ emissions (van der Werf et al. 1993; Ohyama et
al. 2000; Tacconi et al. 1999). The HI absorbing gas samples a
larger volume than the OH, and different structural components may
contribute to the HI and molecular absorption component. However,
the central OH and HI velocity gradients of about 0.25 \kmss
pc$^{-1}$ are surprisingly different from the velocity gradient of
the CO $J$ = 2$-$1 emission of 0.74 \kmss pc$^{-1}$. Possibly the
CO data also samples the higher gradients of the gas in the two
disks.

The uniformly large line widths found in the HI and OH data
confirm large-scale contributions to the central absorption. The
large half-width values for HI and OH of 200 and 75 \kms are
consistent with the stellar velocity dispersion peaking at 270
\kms close to the H$_2$ and CO $J$ = 2$-$1 emission peaks (see
Tecza et al. 2000). The large zero-intensity line widths of HI and
OH confirm the presence of a disturbed and rather clumpy medium
with multiple structures in the central region including the two
nuclei (see also Beswick et al. 2001).

The OH$-$HI optical depth ratio in the central region suggests an
OH/HI abundance ratio of 8.6 x 10$^{-7}$. This value is relatively
high compared to other extragalactic absorption systems (see Baan
et al. 1985), and would support the notion that the central region
consists of enriched disk gas.

\subsection{The two nuclei}

Our data do not provide the detailed properties of the gas in the
nuclear region and in the foreground. The HI and OH column
densities suggest that the more obscured nucleus N1 lies behind
the less obscured northern nucleus.  The path to N1 would then
sample a more complex multiple absorption structure (see also
Beswick et al. 2001). The total HI column density at N1 of 1.28 x
10$^{22}$ cm$^{-2}$ agrees with the column density estimated from
X-ray nuclear emission (Komossa et al. 2003). In addition, the
lower HI column at N2 of 1.01 x 10$^{22}$ cm$^{-2}$ also agrees
with the X-ray column. The ratio of (integrated) OH and HI optical
depths ranges between 0.25 at N1 to 0.32 at N2 , which suggests
lower enrichment at the nuclei relative to the central region.

Both nuclei have AGN characteristics in the radio and X-ray
(Gallimore et al,; Beswick et al. 2001; Komossa et al. 2003).
However, the extended radio continuum seen in the nuclear region
forming the back-ground for the HI and OH absorption is associated
with intense star-formation resulting from the close interaction
of the galaxies (Genzel et al. 1998; Beswick et al. 2001).
Extended radio structures of at least 1 kpc are commonly found in
radio-quiet Seyferts and LINERs that do not follow the morphology
of a galactic disk (Gallimore et al. 2006). Such extended
structures result from nuclear outflows rather than starbursts and
are likely to have a relatively luminous, compact radio source in
the nucleus. Although NGC\,6240 has evidence of a weak nuclear jet
in the northern nucleus N2 (Gallimore \& Beswick 2004), there is
no evidence of strong AGN-related activity that could explain the
total emission region covering a projected 7 x 4 kpc region that
does not include the arc structure. It is more plausible that the
AGNs in N1 and N2 have recently become active as a result of the
interaction, and that the extended radio emission is due to
distributed star-formation and the symmetric outflow triggered by
the star-formation.

The distinct radio continuum region designated N3 is not
necessarily a third nucleus, but is is rather an area of enhanced
(superposed) star-formation region in the interaction zone of the
two galaxies and is currently embedded in the extended radio
structure in the north.

\subsection{The dynamics of NGC\,6240}

The nuclei N1 and N2 are separated by 1.575" corresponding to 793
pc. If the interacting galaxies were in the plane of the sky, they
would be at a very different velocity and the nuclear regions
would be coalesced and extremely confused. Since we see apparently
distinct nuclear entities, they are more distant from each other
and have only a projected distance of 793 pc. Considering that the
highest HI and molecular column densities lie in between the
nuclei and that the X-ray source in N1 shows the highest column
density, it is most plausible that N1 lies behind N2. The small
velocity difference suggests that the N1-N2 connecting axis has a
small angle with the line-of-sight and the relative values of the
velocities at N1 and N2 suggest that the galaxies are just past
transit.

The HI/OH systemic velocities at the nuclei can be used for a
dynamical/orbital scenario for the nuclei projected on the sky.
Given the relatively large beam, we find an HI estimate of 7235
and 7305 \kms and an OH estimate of 7255 and 7370 \kmss, which are
nominally consistent with the higher resolution HI values of 7258
and 7440 \kms (Beswick et al. 2001). These values are larger than
the difference of stellar velocities of 50 \kms (Tecza et al.
2000), but consistent with the H$_2$ ($\approx$ 150 \kmss),
CO(2$-$1) ($\approx$ 100 \kmss), and Bracket $\gamma$ emission
data (Lira et al. 2002; Ohyama et al 2000; Tecza et al. 2000).  We
adopt the high-resolution HI estimate of the velocity difference
for the nuclei $\delta V$ = 182 \kmss. The projected distance
$D_{obs}$ of the connecting line between the two galaxies is
1.575", which corresponds to a projected separation of 793 pc.

As an attempt at the dynamics of the close encounter, we assume a
simple {\it edge-on circular orbit for two equal masses M} around
the center of mass, and a (small) projection angle $\theta$
between our line of sight and the connecting line between the two
nuclei. The description of the orbital motion of the system
follows from: sin($\theta$)$^3$ = 0.031 $\delta V^2$ $G^{-1}$
$M_n^{-1}$ $D_{obs}$, where M$_n$ is the combined dynamic mass of
the nuclear region. The estimate of Tecza et al. (2000) of the
stellar mass in each of the nuclei of about 2 x 10$^9$ \msols
gives a combined dynamic mass of the nuclei of $M_n$ = 1.2 x
10$^{10}$ \msol using the smaller velocity difference. The central
gas concentration also constitutes a significant fraction of the
dynamic mass, such that $M_{gas}(R\leq 470 pc)$ $\approx$ (2$-$4)
x 10$^9$ \msol $\approx$ (0.3$-$0.7) $M_{dyn}$ (Tacconi et al.
1999).  For this reason, we adopt a dynamic mass for the nuclei of
$M_n$ = 1.5 x 10$^{10}$ \msols, which results in $\theta$ =
13.4$^o$ (projection factor = 4.3), an orbital velocity of 392
\kmss, and a distance between the nuclei of 3.42 kpc. The orbital
period is about 27 Myr. This scenario gives a sufficiently large
separation distance to ensure identifiable nuclear/galaxy
characteristics at this well-advanced stage just before
coalescence.

\subsection{The two interacting galaxies}

The OH and HI velocity field of the nuclear region is dominated by
the large-scale organized motion of the accumulated gas structure.
However, the detailed velocity pattern displayed in Figures
\ref{himom12}a and \ref{ohmom12}a suggest the presence of
large-scale velocity components associated with the interaction of
the two galaxies. The analysis of the stellar velocity field in
the southern galaxy suggests a northwest-southeast rotation (i =
60$^o$) at PA = $-$34$^o$ with $V_{rot}$ = 270 $\pm$ 90 \kms
(Tecza et al. 2000). The northern galaxy (i = 33$^o$) displays a
southwest-northeast rotation at PA = 41$^o$ with $V_{rot}$ = 360
$\pm$ 195 \kmss.

The HI and OH absorption in the northeast extension (Figs.
\ref{himom12}a and \ref{ohmom12}a) shows a southwest-northeast
rotation for the northern galaxy at PA = 40$-$50$^o$ with an HI
gradient of 1.15 (or 0.75 for OH) \kmss pc$^{-1}$ north of N2,
which is consistent with a stellar gradient of 1.2 \kmss
pc$^{-1}$. However, the region south of N1 also shows evidence of
a south-north rotation at 1.49 \kmss pc$^{-1}$ for HI and 0.75
\kmss pc$^{-1}$ for OH, and is associated with the large gas
structure of the interaction. In addition, there is a weak
low-velocity HI signature south of N1 down to 6900 \kms with the
southwest-northeast gradient of 0.89 \kmss pc$^{-1}$, which could
constitute the motion of the southern galaxy. The absence of a
clear velocity signature of the southern galaxy could easily
result from the column density around N1. Furthermore, the region
west-south-west of N1 is very perturbed by the outflow and shows
evidence of components with velocities up to 7750 \kms.

\subsection{The superwind outflow}

The continuum structure extends from N1 and N2 in all directions,
including to the northwest region and the radio arc. The western
radio arc results from the shocked regions forming the boundaries
of a symmetric superwind-driven outflow emanating from N1 (see
Heckman et al. 1990), that is less prominent towards the east. In
addition, there is a considerable extended radio emission
resulting from distributed star-formation in the southwest and
northeast regions. Besides the presence of two AGNs, the radio
properties of the nuclear region suggest dominant starburst
activity (Beswick et al. 2001; Gallimore \& Beswick 2004). The
K-band emission at both nuclei suggests a dominant population of
red supergiants (Tecza et al. 2000). Recent X-ray data suggest
that the outflow was indeed symmetric and that there are remnants
on both sides of the nuclei.

The HI data shows a blue-shifted component along the line of sight
extending to $-$300 \kms with respect to N1 (Fig. \ref{hipvr}).
Similarly there is a low velocity component in the OH data at
$-$120 \kms (Fig. \ref{ohpv}), which produces a wing on the 1667
MHz line (Fig. \ref{ohspec}). Besides Ohyama et al. (2000) note a
$-$250 \kms component in the H$_2$ emission. These blue-shifted
features could represent line-of-sight outflows and shocks, which
are driven into the denser nuclear ISM by the nuclear starburst
and are associated with the superwind.

The HI and OH velocity and linewidth data west-south-west of N1 as
well as the blue-shifted HI components at N1 clearly confirm that
N1 is the origin of the outflows. The lifetime of the starburst
has been estimated at $\approx$ 10 Myr (Tecza et al. 2000), which
is about 40\% of the orbital period of 27 Myr derived above. The
orbital motion may thus have resulted in smearing out the X-ray
and radio emission regions. In addition, the difference of the
emission strength of the northern and southern parts of the radio
arc may also have resulted from the "piling up" of emission in the
forward direction, which confirms that N1 is moving north. The
complicated OH and HI structures southwest of N1 are associated
with the outflow into the western cavity and cover a large
velocity range, with a mean of about 100 \kms below that of the
systemic velocity of N1. There is (marginal) evidence of HI
absorption components west-south-west of N1 reaching an extreme of
7750 \kmss. Our HI data also displays continuous absorption
against the base of the northern radio arc and against arc
components W0, W2 and W4.

\subsection{The extended absorption}

The HI absorption is found across much of the extended radio
emission, while the OH is found only in the central region of the
source. The HI absorption shows complicated structures with a wide
range of velocities and line widths, and relatively low column
densities. Some of this material is associated with foreground
dust lanes and ejected gas resulting from the interaction.

The extended radio emission in NGC$\,$6240 is associated with the
remnants of the two galaxies and the radio arcs resulting from the
symmetric superwind outflows emanating from N1 and possibly N2. As
discussed above, we find significant absorption and an HI velocity
gradient of 1.0 \kmss pc$^{-1}$ in the NE region, which is
associated with the remnant of the northern galaxy. In addition,
there is an east-to-west gradient towards the NW region (at PA =
-50$^o$), where we find the highest HI velocities in the system.

The column density map of Figure \ref{hicolumn} displays
widespread HI absorption in the large-scale structure of
NGC$\,$6240. Discrete absorption components with column densities
in the range of 0.4 - 1.0 x 10$^{22}$ cm$^{-2}$ are found at the
continuum component in the NW region and at the W0, W2, and W4
components of the arc structure. The velocities at these
components, which are not associated with distinct components in
the optical and H$\alpha$, suggest an increasing velocity towards
the south.

The dominant absorption at W0 is caused by a north-south dust-lane
passing in the foreground of the continuum structure with an
estimated column density of 3.2 x 10$^{21}$ cm$^{-2}$. Images with
various optical and X-ray instruments (Max et al. 2005; Komossa et
al. 2003) show the clear presence of this north-south dust-lane
that crosses the NW radio structure at W0 and accounts for an
added column density. While the distributed HI in the NW region
has a velocity of about 7500 \kmss, the dust-lane has a systemic
velocity of about 7270 \kmss, which is close to that of N1.
Furthermore, it has no clear velocity gradient because of its
distance from the nuclei.

The radio continuum and X-ray images also display two extended
structures S and SW of nucleus N1 (see Fig \ref{continuum}b;
Komossa et al. 2003). It is found that a dust-lane structure
towards the south divides these two structures. Only weak
absorption has been seen against the continuum in this southern
region.

\subsection{The OH and H$_2$O emission}

A narrow H$_2$O maser line has been detected towards the nuclear
region of NGC$\,$6240 at Vlsr = 7565 \kms located within 3 pc from
the continuum peak at N1 (Hagiwara et al. 2003). The occurrence of
an H$_2$O maser is rather unusual in a FIR dominated galaxy such
as NGC$\,$6240, which is more likely an OH Megamaser candidate.
OH-MM UGC\,05101 also hosts a weak H$_2$O maser towards the
nuclear region (Zhang et al. 2005). The maser in NGC\,6240 is
redshifted about 300 \kms relative to the systemic velocity of N1,
while high-velocity OH or HI gas has only been found in the
northern region of the source (Figs. \ref{hipvd}, \ref{hipvr}, and
\ref{ohpv}). An association of the maser with the AGN could exist
with shocked outflows, or jet-molecular cloud interactions in
order to account for these discrepant velocities. Examples of
other redshifted jet-related masers can be found in the elliptical
NGC\,1052 (100-180 \kms; Claussen et al. 1998) and Mkn\,348 (130
\kmss; Peck et al 2003). Alternatively, there could be either an
association with infalling foreground gas to the nuclear region or
with an active nucleus.

The shallow and multi-component Arecibo spectrum of OH in
NGC$\,$6240 has been interpreted on the basis of partial infilling
of the absorption by emission (Baan et al. 1985). The asymmetries
in the spectrum of Fig. \ref{ohspec} and the PV diagram Fig.
\ref{ohpv} could indeed support this notion. While asymmetries
suggest emission infilling at the velocity of N2, there is no
evidence in the data for this. The bulk of the OH absorption shows
an LTE line ratio. Only the western side of the central absorption
shows non-LTE ratios that suggest infilling with emission at the
low-velocity side of the 1667 MHz line at the velocity of N1.
Non-LTE conditions could be caused by the FIR radiation field,
which is dominant in NGC\,6240 and has the right infrared colors
for FIR pumping as in OH Megamasers (Baan 1989; Henkel and Wilson
1990). While there would be enough background radio continuum for
this purpose in the N1 system, there is no discernable line
emission.

\section{Summary}

The extended HI and OH absorption against the continuum structure
has revealed more of the dynamic and evolutionary properties of
the interacting system NGC\,6240, and complementary evidence
obtained at other wavelengths. The radio continuum structure and
the associated absorption structure of NGC\,6240 is in part the
result of a superposition of the two galaxies and their
constituents. In a simple dynamic model using HI systemic
velocities, the northern galaxy with nucleus N2 would be located
in front of the southern galaxy with nucleus N1, such that the
N1-N2 connecting line would be foreshortened by a factor of 4.3.
In this picture N1 would be expected to have the largest absorbing
column density, while the central disks of the galaxies would be
superposed between the two nuclei.

The radio continuum structure of 15" x 17" (7.6 x 8.6 kpc) peaks
at the two nuclei of the interacting galaxies with hybrid
starburst and AGN emission, and is surrounded by an extended
structure associated with star-formation activity triggered by the
interaction. A large-scale but incomplete loop structure on the
western side of the source has been associated with a nuclear
blowout and outflow from nucleus N1 of the southern galaxy, while
traces of a similar structure can also be found at the eastern
side of the source.

The HI absorption covers a contiguous 5" x 8" (2.7 x 4.2 kpc)
region of the continuum structure and provides a large-scale view
of the velocity field across this area. The peak of the HI
absorption falls close to nucleus N1 in such a way that the HI
column density at N1 is 1.26 times that of N2, and is in agreement
with the estimates from X-ray observations.

OH absorption has been found only against the nuclear continuum
and extends 2.5" x 2.0" (1.25 x 1.0 kpc). The largest column
density of the OH absorption falls north of N1 and about halfway
between N1 and N2, a fact that agrees with maps of other molecular
emissions. The column densities at N1 and N2 are about 60\% of
that of the central gas structure.

The HI and OH velocity fields reveal parts of the velocity
gradients of the two individual galaxies buried in the central
region. Velocity gradients at various locations suggest gas
motions resulting from the interaction of the two galaxies. In
particular, the location of N1 and the region to the west displays
blue-shifted (l.o.s.) outflow components, as well as structural
components related to the sideways outflow into the western
bubble. This evidence clearly confirms the nuclear activity at N1
as the origin of the outflows and the cause of the radio arc,
which is consistent with the evidence from X-ray and H$\alpha$
data.

Distinct velocity components are found in the northern region and
along the radio structure northwest of the nuclei. A foreground
dust lane passes across the northwest radio loop structure.
Absorption in more distant continuum components do not reveal a
coherent velocity pattern. The large width of the HI absorption
line across the central part of the source confirms the violent
dynamics of the system. The highest OH velocity widths are found
at the central gas deposit, but they are significantly lower than
those of HI and therefore less affected by the merger dynamics.

The central gas structure may result from a superposition of the
disks of the two galaxies. There may also be accumulation of gas
in the center of mass of the dynamic system. The central gas
accumulation between the nuclei behaves as an independent
structure with a velocity gradient proportional to the velocity
difference of the two nuclei, and the gas appears locked into the
motion of the system. The HI and OH velocity gradients for the
central region are much smaller than that of CO(2$-$1), which may
suggest that different observations detect distinctly different
scale sizes within these structures.

The OH hyperfine ratio in the absorption region suggests mostly
LTE conditions across the nuclear region, except in the western
part of the central gas accumulation where non-LTE conditions are
found. Non-LTE conditions may suggest that radiative far-infrared
pumping actively reduces the absorption on the low-velocity side
of the 1667 MHz OH line. No further OH maser emission has been
found in the system.

%
%

\acknowledgements
WAB would like to thank Aubrey Haschick (formerly of Haystack
Observatory) for constructive support during the early stages of
this project.

%
%

%
\def\apj{ApJ }

\begin{deluxetable}{lcccccc}
\tabletypesize{\footnotesize} \tablecolumns{11} \tablewidth{0pt}
\tablecaption{Continuum Fluxes} \tablehead{ \colhead{} &
\colhead{} & \colhead{} & \colhead{} & \colhead{1420 MHz} &
\colhead{} & \colhead{1667 MHz} \\
\colhead{Component} & \colhead{$\alpha$(B1950)} &
\colhead{$\delta$(B1950)} & \colhead{I$_{\nu}$} &
\colhead{S$_{\nu}$} & \colhead{I$_{\nu}$} &
\colhead{S$_{\nu}$} \\
\colhead{} & \colhead{h m s} &
\colhead{\arcdeg\phn\arcmin\phn\arcsec} &
\colhead{mJy$\,$b$^{-1}$} & \colhead{mJy} &
\colhead{mJy$\,$b$^{-1}$} & \colhead{mJy} \\} \startdata
N1     & 16 50 27.84 & 02 28 57.5 & 108.8&225.9 & 55.7& 88.2 \\
N2 + N3& 16 50 27.84 & 02 28 58.7 &      &      & 41.4& 74.1 \\
NW     & 16 50 25.70 & 02 29 00.2 & 6.03 &    -  & 3.3 &  6.3 \\
W0     & 16 50 27.43 & 02 29 02.3 & 15.7 & 15.5 & 6.1 & 24.0 \\
W1     & 16 50 27.18 & 02 29 02.6 & 6.65 &    -  & 3.4 &  6.4 \\
W2     & 16 50 27.16 & 02 29 00.2 &  13.8& 14.1 & 5.2 & 16.4 \\
W3     & 16 50 26.92 & 02 28 56.6 &  11.1&   -   & 4.4 & 14.4 \\
W4     & 16 50 27.20 & 02 28 55.1 &  9.8 &  9.7 & 4.4 & 12.1 \\
S      & 16 50 27.78 & 02 28 53.3 &  14.6& 13.8 & 6.2 & 17.9 \\
E1     & 16 50 28.20 & 02 28 57.1 &  1.5 &  -   &  -   &  -   \\
NE     & 16 50 28.08 & 02 28 59.0 &  1.8 &  -   &  -   &  -   \\
SE     & 16 50 27.90 & 02 28 55.1 &  4.0 & 5.7  &  -   &  -   \\
\enddata
\tablecomments{The 21\,cm and 18\,cm continuum maps have an rms of
respectively 0.30 and 0.27 mJy\,beam$^{-1}$.}
\end{deluxetable}

\footnotesize
\begin{deluxetable}{lcccccccccc}
 \tabletypesize{\scriptsize} \tablecolumns{11}
\tabletypesize{\footnotesize} \tablewidth{0pt} \tablecaption{HI
and OH Absorption Parameters} \tablehead{ \colhead{Component} &
\colhead{V(HI)\tablenotemark{a}} &
\colhead{$\Delta$V\tablenotemark{b}} & \colhead{$\tau_{HI}$} &
\colhead{$\int\tau_{HI}dv$} & \colhead{N$_H$\tablenotemark{c}} &
\colhead{V(OH67)\tablenotemark{a}} &
\colhead{$\Delta$V\tablenotemark{b}} & \colhead{$\tau_{OH67}$} &
\colhead{$\int\tau_{OH}dv$} &
\colhead{N$_{OH67}$\tablenotemark{d}} \\
\colhead{} & \colhead{(km/s)} & \colhead{(km/s)}
&&\colhead{(km/s)} &\colhead{(cm$^{-2}$)} & \colhead{(km/s)} &
\colhead{(km/s)} && \colhead{(km/s)}& \colhead{(cm$^{-2}$)} }
\startdata Nucleus N1 & 7295& 909& 0.15 & 70.4$\pm 0.7$ & 1.28
(22)& 7243& 295 & 0.038 & 1.39$\pm 0.1$  & 6.55 (15)\\
Central Peak & 7295& 913& 0.12 & 70.4$\pm 0.7$ & 1.28 (22)& 7274&
406 & 0.063 & 2.29$\pm 0.2$ & 1.1 (16)\\
Nucleus N2   & 7339& 870& 0.11 & 55.5$\pm 0.5$ & 1.01 (22)& 7363&
406 & 0.035 & 1.28$\pm 0.1$  & 6.03 (15)\\
NE corner    & 7600& 303& 0.52 & 25.6$\pm 1.5$ & 4.67 (21)& - & - & - & & - \\
NW comp & 7513& 606& 0.07 & 34.2$\pm 1.3$ & 6.23 (21)& - & - & - & & - \\
Dust lane    & 7270& 800& 0.05 & 51.7$\pm 1.2$ & 9.42 (21)& - & - & - & & - \\
E1 comp & 7530& 380& 0.04 & 46.8$\pm 1.3$ & 8.55 (21)& - & - & - & & - \\
W2 comp & 7034& 217& 0.04 & 21.4$\pm 1.5$ & 3.90 (21)& - & - & - & & - \\
W4 comp & 7687& 390& 0.07 & 42.7$\pm 1.3$ & 7.80 (21)& - & - & - & & - \\
\enddata
\tablenotetext{a}{Peak values.} \tablenotetext{b}{Full width
zero-intensity values.} \tablenotetext{c}{Assumed T$_S$=100 K for
HI.} \tablenotetext{d}{Assumed T$_{ex}$=20 K for OH.}
\end{deluxetable}

\end{document}